\def\year{2018}\relax
\documentclass[letterpaper]{article} %DO NOT CHANGE THIS
\usepackage{aaai}  %Required
\usepackage{times}  %Required
\usepackage{helvet}  %Required
\usepackage{courier}  %Required
\usepackage{url}  %Required
\usepackage{graphicx}  %Required
\usepackage{bm}
\usepackage{color}
\usepackage{amssymb}
\usepackage{ulem}
\definecolor{red}{rgb}{1,0,0}

\begin{document}

\title{Molecular Flow Monte Carlo}
\author{
Katsuhiro Endo, Daisuke Yuhara and Kenji Yasuoka\\
Department of Mechanical Engineering, Keio University\\
3-14-1 Hiyoshi, Kohoku-ku, Yokohama\\
Kanagawa, Japan 223-8522\\
}

\maketitle
\begin{abstract}
In this paper, we suggest a novel sampling method for Monte Carlo molecular simulations. 
In order to perform efficient sampling of molecular systems, it is advantageous to avoid extremely high energy configurations while also retaining the ability to quickly generate new and independent trial states. 
Thus, we introduce a continuous normalizing flow method which can quickly generate independent states for various proposal distributions using a first-order differential equation. 
We define this continuous normalizing molecular flow approach based on two-body intermolecular interactions to achieve a probability distribution transformation method which yields distributions which have probability densities of zero when molecule pairs are in close proximity; while in all other cases, the probability density is compressed such that it is spatial uniform. 
This transform provides the proposal distribution which generates no states of extremely high energy.
We find that an inverse square flow is applicable as the continuous normalizing molecular flow. 
Using the transformed distribution, we can perform the Metropolis-Hastings method more efficiently.
The high efficiency of the proposed method is demonstrated using simple molecular systems.
\end{abstract}

\section{1. Introduction}
Monte Carlo molecular simulation is one of the most commonly used and powerful tools with which to investigate the physical properties of molecular systems. It has seen widespread use in various fields, e.g., polymers \cite{Escobedo1996}, proteins \cite{Andrzej1994}, nucleation \cite{Mandell1976}, and hydrates \cite{Mezei1983}. The method works by generating states that follow the Boltzmann distribution for specific molecular systems, for example,
\begin{eqnarray}
  		p({\bm r}^N) \propto \exp{\{-\beta U({\bm r}^N) \}} 		\label{eq11},
		\\ {\rm where} \ U({\bm r}^N) = \sum_{i < j}{ U_{ij}(\|{\bm r}_i-{\bm r}_j\|) },
\end{eqnarray}
${\bm r}^N=({\bm r}_1,{\bm r}_2,...,{\bm r}_N)$ denotes the total set of 3-dimensional $N$ molecule coordinates, $\beta$ is the inverse temperature, $U({\bm r}^N)$ is the total potential energy of system, and $U_{ij}$ is the two-body intermolecular potential energy between molecules $i$ and $j$.
To sample the Boltzmann distribution, we usually use dedicated Monte Carlo methods such as the Metropolis-Hastings algorithm \cite{Hastings1970}, replica exchange method \cite{Sugita1999}, and umbrella sampling \cite{Torrie1977}, according to the characteristics of the system being studied.
Despite the applicability of the Monte Carlo approach, performing large-scale simulations is not practical because existing Monte Carlo methods generate new states which are quite similar to preceding ones. The implication of this is that, for large systems, it takes a long time to generate a sufficient number of independent states, which are required to obtain unbiased results. 
There is, therefore, a need for fast and efficient methods with which to generate independent states for large systems.

Recently, a notable sampling method, continuous normalizing flow \cite{Ricky2016}, was proposed. This method quickly generates independent states which follow various distributions using a parametric first-order differential equation. Additionally, the probability density of each state may also be calculated. The method is suitable for use with Monte Carlo simulations and has great potential to improve sampling efficiency. 

In this paper, we examine the applicability of continuous normalizing flow for molecular systems.
We define continuous normalizing molecular flow based on two-body intermolecular interactions.
Furthermore, we find that inverse square flow is suitable for continuous normalizing molecular flow.
This flow is a space-constant instantaneous change flow in 3-dimensional space, which can be used even in the presence of periodic boundary conditions. Using this flow as a sampler for proposal distributions, we present a highly efficient Metropolis-Hastings method, named Molecular Flow Metropolis-Hastings (MFMH). The efficiency of MFMH is evaluated through simulations of simple molecular systems.

%Space-constant instantaneous change flow is required to sample larger systems ensembles is also showed by numerical calculation. Moreover, the simulations show that the proposed method generates states very efficiently.

\section{2. Continuous normalizing molecular flow }
Continuous normalizing flow generates the state ${\bm z}(t)$ by solving the first-order differential equation
\begin{eqnarray}
  		\frac{d{\bm z}}{dt}=f({\bm z}(t),t),
\end{eqnarray}
where $f$ is a uniform Lipschitz continuous function in ${\bm z}$, while also being continuous in $t$.
The distribution $p({\bm z}(t))$ that state ${\bm z}(t)$ follows is easily calculated by solving following differential equation
\begin{eqnarray}
  		\frac{\partial {\rm log} \ p({\bm z}(t)) }{\partial t}= 
		-{\rm tr}\left( \frac{df}{d{\bm z}(t)} \right)
\end{eqnarray}
from the initial distribution $p({\bm z}(0))$. The term ${\rm tr}(df/d{\bm z})$ is referred to as the instantaneous change of variables.
Here, let us consider ${\bm z}$ as the total set of molecular coordinates ${\bm r}^N=({\bm r}_1,{\bm r}_2,...,{\bm r}_N)$ and $f$ as two-body intermolecular ``force''. 
Solving first-order the differential equation gives
\begin{eqnarray}
  		\frac{d{\bm r}_i}{dt}= \sum_{i \neq j}{ f_{ij}( \|{\bm r}_i-{\bm r}_j\|) ({\bm r}_i-{\bm r}_j}) \
\end{eqnarray}
for all $i$, where $f_{ij}(r)$ is the two-body intermolecular function. In this case, the instantaneous change of ${\bm r}^N$ is described as 
%  		\frac{\partial {\rm log} \ p({\bm r}^N(t)) }{\partial t}= 
\begin{eqnarray}
		{\rm tr}\left( \frac{df}{d{\bm r}^N(t)} \right) = 
		\sum_{i,j}  Df_{ij}(r_{ij}) + r_{ij}\frac{df_{ij}}{dr_{ij}}(r_{ij}),
\label{eq21}
\end{eqnarray}
where $r_{ij} =  \|{\bm r}_i(t)-{\bm r}_j(t)\|$, and $D$ is the number of spatial dimensions of the system (usually $D=3$).
Note that if we regard continuous normalizing molecular flow as machine learning architecture, then this architecture has novel properties: (particle) permutation invariance and translational/rotational invariance in the presence of periodic boundaries.

\section{3. Inverse square flow}

\subsection{Inverse square flow on no boundary condition } 
\begin{figure}[tbp]
\begin{center}
\includegraphics[width=80 mm ]{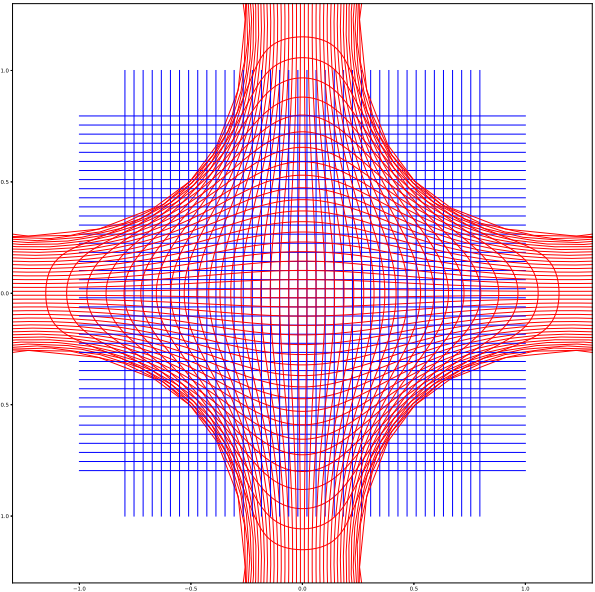}
\caption{
{\bf A distribution which has been transformed by 2-dimensional inverse square flow from an initially uniform distribution. } 
Using the inverse square flow process, molecules are remapped by those molecules which have been fixed in each of the four corners of the plot. This figure shows how molecules which are placed on the uniform blue grid are then remapped onto the red grid.} %The area of each grid remains constant through the inverse square flow process. Furthermore, the occurrence of molecule pairs in close proximity is prevented in the new distribution.}
\label{fig31}
\end{center}
\end{figure}

In this section, we introduce the inverse square flow which achieves a space-constant instantaneous change of ${\bm r}^N$ while preventing the generation of states with molecule pairs in close proximity.
By solving the equation $Df+rdf/dr=0$, to satisfy the condition that the instantaneous change of ${\bm r}^N$ (Eq.\ref{eq21}) must always be zero, we obtain the following solution:
\begin{eqnarray}
		f_{ij}(r) = \frac{1}{r^{D}}.
\end{eqnarray}
We may call this solution $D$-dimensional inverse square flow because the term ${\bm r}/r^D$ takes the same form as the inverse-square law.
In Fig.\ref{fig31} we present the numerical results of a 2-dimensional inverse square flow.
We see the distribution of a molecule transformed from a uniform distribution in response to the inverse square force from immobile molecules placed at the four corners of the plotted area.
In the transformed distribution, we can confirm that the occurrence of molecule pairs positioned in close proximity is prevented, while the probability density is the same as the untransformed distribution since the instantaneous change of ${\bm r}^N$ is 0.
Therefore, by using inverse square flow, we can create a transformation that does not change the probability density and prevents the generation of states where molecules are in close proximity. 
In Monte Carlo molecular simulation, any state ${\bm r}^N$ which includes molecule pairs in close proximity will always have a very high energy $U({\bm r}^N)$. It is computationally inefficient to attempt the sampling of such states.
Thus, the method presented here has the potential to greatly enhance the efficiency of sampling within Monte Carlo simulations. This promises to be especially useful for systems with strongly repulsive potentials and/or high-density phases.

%p({\bm r}^N) \propto \exp{\{-\beta U({\bm r}^N) \}}
%excludes too close molecule pairs
%This is because of the curse of dimensionality
%Even if the term $Df_{ij} + rdf_{ij}/dr$ is small, the sum $\sum_{i,j}$ is enlarged quadratically. Thus, 
%次元の呪いについての図解を入れる????

\begin{figure}[tbp]
\begin{center}
\includegraphics[width=60 mm ]{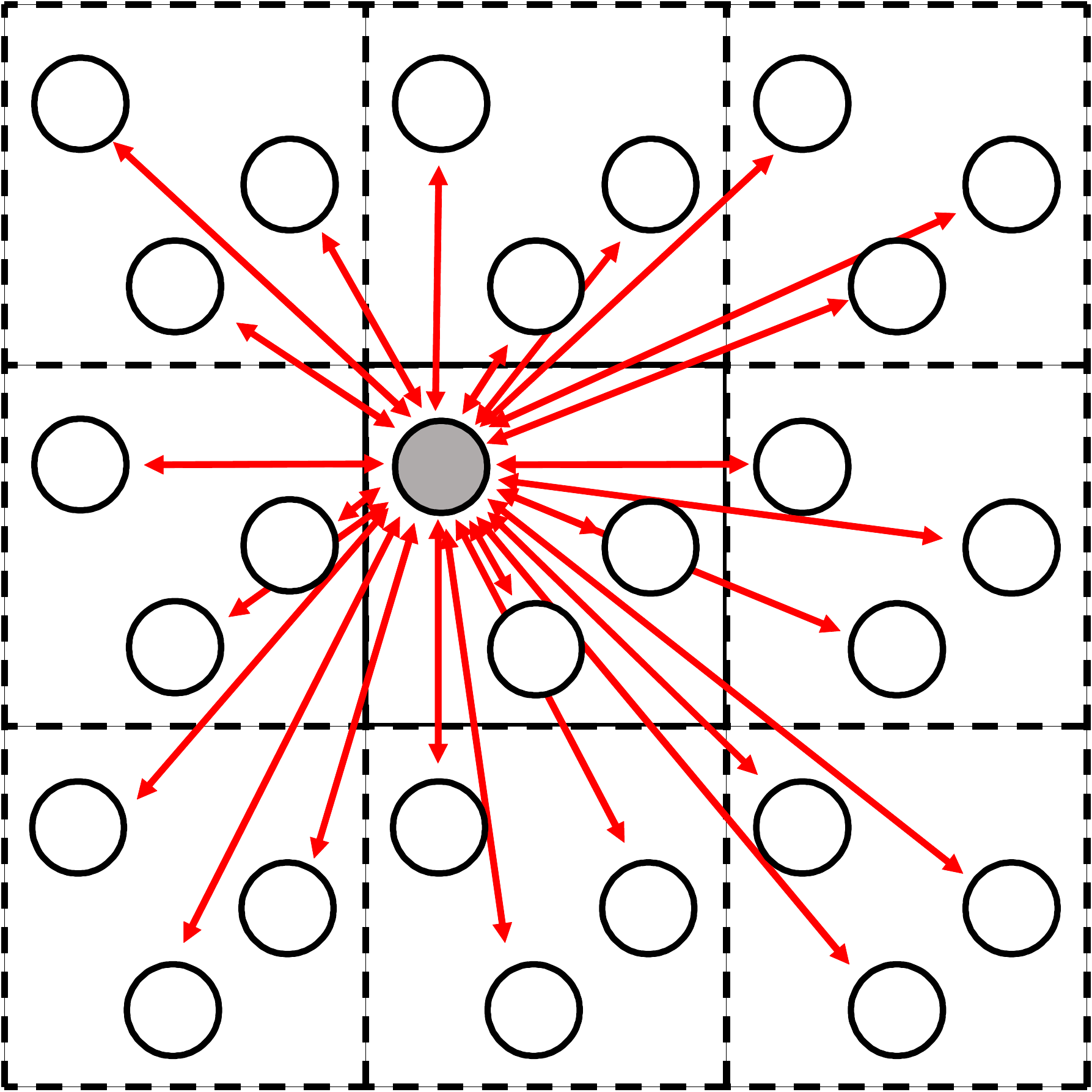}
\caption{
{\bf Periodic boundary conditions and periodic molecule images. } 
With periodic boundary conditions, the simulated system cell is copied multiple times to create periodic images which are positioned such that the simulated cell is surrounded in all directions, including the diagonals. Therefore, it is not only those molecules in the central cell which are affected by the movement of molecules, but also those in the periodic images. The red arrows show those molecule pairs which affect the movement of the grey molecule.
}
\label{fig32}
\end{center}
\end{figure}

\subsection{Inverse square flow with periodic boundary conditions }
We perform molecular simulations in the presence of periodic boundary conditions. With periodic boundary conditions, each molecule is affected by other molecules in the original cell and the images of molecules in the neighbouring periodic cells that surround the simulated cell (see Fig.\ref{fig32}). For simplicity, let us consider that a cell is a cube with each side being a unit in length. 
Continuous normalizing molecular flow for periodic boundary conditions is described as 
\begin{eqnarray}
  		\frac{d{\bm r}_i}{dt}= \sum_{\bm n}\sum_{i \neq j}{ f_{ij}( \|{\bm r}_i-{\bm r}_j-{\bm n}\|) ({\bm r}_i-{\bm r}_j-{\bm n}})
\label{eq31}
\end{eqnarray}
and the instantaneous change of ${\bm r}^N$ is described as
\begin{eqnarray}
		{\rm tr}\left( \frac{df}{d{\bm r}^N(t)} \right) = 
		\sum_{\bm n} \sum_{i \neq j}  Df_{ij}(r_{ij,{\bm n}}) + r_{ij,{\bm n}}\frac{df_{ij}}{dr_{ij,{\bm n}}}(r_{ij,{\bm n}}),
\end{eqnarray}
where $r_{ij,{\bm n}}=\|{\bm r}_i-{\bm r}_j-{\bm n}\|$ and ${\bm n} \in \mathbb{Z}^3$. 
Unfortunately, when simply applying inverse square flow to systems with periodic boundary conditions, the sum of all periodic forces (the righthand term in Eq.\ref{eq31}) does not converge because the inverse square force is a long-range force. To handle inverse square flow, we use Ewald summation \cite{Ewald1921} for the calculation of this force when $D=3$. In Ewald summation, the inverse square force is divided into
\begin{eqnarray}
\sum_{\bm n}{ \|{\bm r}-{\bm n}\|^{-3} ({\bm r}-{\bm n})} 
= \sum_{\bm n}{\bm \nabla}_{\bm r} \frac{1}{\|{\bm r}-{\bm n}\|} = \nonumber \\
{\bm \nabla}_{\bm r} \left( \sum_{\bm n}
 \frac{{\rm erfc}(G\|{\bm r}-{\bm n}\|)}{\|{\bm r}-{\bm n}\|}
+\frac{1-{\rm erfc}(G\|{\bm r}-{\bm n}\|)}{\|{\bm r}-{\bm n}\|}
\right).
\label{eq32}
\end{eqnarray}
The left term and right term in Eq.\ref{eq32} are summed in real space and reciprocal space, respectively. $G$ is the screening factor that determines the relative proportion of the real and reciprocal space sums. 
Using a pairwise form of the Ewald sum \cite{Yi2017}, instead of direct calculation of the periodic sum, gives
\begin{eqnarray}
\sum_{\bm n} \|{\bm r}-{\bm n}\|^{-3} ({\bm r}-{\bm n}) = \nonumber
\\
{\bm \nabla}_{\bm r} \left(
\sum_{\bm n} { \frac{{\rm erfc}(G\|{\bm r}-{\bm n}\|)}{\|{\bm r}-{\bm n}\|  }} 
+
4\pi \sum_{\bm k \neq 0}{
\frac{ e^{-\|\bm k\|^2/4G^2} e^{i{\bm k}\cdot {\bm r}}  }{\|\bm k\|^2}
}
\right),
\label{eq33}
\end{eqnarray}
where ${\bm k} \in { 2\pi\mathbb{Z}^3 }$ is the reciprocal vector. This leads to converge for any $G \in (0,\infty)$. A non-derivative form of the pairwise force is given in Appendix A. 
By expressing Eq.\ref{eq33} as ${\bm f}^p({\bm r})$, the inverse square flow for periodic boundary conditions may be written in a more compact form:
\begin{eqnarray}
  		\frac{d{\bm r}_i}{dt}= \sum_{i \neq j}{ {\bm f}^p({\bm r}_i-{\bm r}_j) }.
\end{eqnarray}
Note that this equation does not have the problem of non-convergence of the periodic sum.
Notably, though we only use the inverse square flow, the periodic sum of the inverse square flow has a non-zero instantaneous change. The instantaneous change of the periodic sum of the inverse square flow is also space-constant, but the value is $\sum_{i\neq j}4\pi$ (a proof of this is given in Appendix B).

\section{4. Molecular Flow Monte Carlo}

\subsection{Standard Metropolis-Hastings method }
Let us consider the case where $p({\bm r}^N)$ is the target distribution, such as in Eq.\ref{eq11} when states from $p({\bm r}^N)$ are required. Although the mathematical expression of $p({\bm r}^N)$ is usually well-known, the direct generation of states from $p({\bm r}^N)$ is typically difficult. The Metropolis-Hastings (MH) method generates states by stochastically accepting states from a Markov chain. If we iteratively generate states from the Markov chain $q({\bm r}^N_*|{\bm r}^N)$ (the proposal distribution) and accept states with the following probability:
\begin{eqnarray}
 A({\bm r}^N_*,{\bm r}^N) = \min\left(
  1, \frac{p({\bm r}^N_*)}{p({\bm r}^N)} \frac{q({\bm r}^N|{\bm r}^N_*)}{q({\bm r}^N_*|{\bm r}^N)}
 \right),
\end{eqnarray}
then the generated states will follow $p({\bm r}^N)$ \cite{Hastings1970}. In the MH method, the mathematical expression of $q({\bm r}^N_*|{\bm r}^N)$ is required for the calculation of the acceptance probability. Or, if the proposal distribution is symmetric ($q({\bm r}^N_*|{\bm r}^N) = q({\bm r}^N|{\bm r}^N_*) $), the acceptance probability $A({\bm r}^N_*,{\bm r}^N)$ reduces to $ \min\left( 1, p({\bm r}^N_*)/p({\bm r}^N) \right) $, thus, under these conditions, a mathematical expression for $q({\bm r}^N_*|{\bm r}^N)$ is not actually required in practice.

In Monte Carlo molecular simulation, we often move only one molecule at a time. The movement of many molecules often results in a large energy change, and the acceptance probability becomes exponentially small. To maintain a sufficiently high acceptance ratio, we only move one randomly selected molecule at a time.

The most commonly used proposal distribution for $q({\bm r}_{i*}|{\bm r}_i)$ is a normal distribution:
\begin{eqnarray}
q({\bm r}_{i*}|{\bm r}_i)=\frac {1}{(\sqrt{2\pi}{r_{move}})^3}\exp\biggl\{ -\frac {\|{\bm r}_{i*}-{\bm r}_i\|^2}{2{r_{move}}^2}\biggr\},
\label{eq41}
\end{eqnarray}
where ${r_{move}} \in \mathbb{R}^+$ controls molecule movement range and has a symmetric distribution.
The movement size parameter ${r_{move}}$ affects the efficiency of sampling. If ${r_{move}}$ is too small, the acceptance ratio (which is the ratio of the number of proposed states from $q$ to the number of accepted states) becomes very high. However, in this case, the correlation between accepted states is strong. Thus independent states cannot be attained in an efficient manner. On the other hand, if ${r_{move}}$ is too large, independent states are easily sampled but proposal states are rarely accepted.

The procedure of standard MH is as follows:
\begin{enumerate}
	\item Set sample steps $s = 0$. 
	\item Set initial molecule positions ${\bm r}^N(s)$.
	\item Calculate probability $p_{pre} = p({\bm r}^N(s=0))$.
	\item Repeat the process process below until a sufficient number of states have been accepted.
 	\item Generate state ${\bm r}^N_*$ from the proposal distribution $q$ and compare to the previously accepted state ${\bm r}^N(s)$.
 	\item Calculate the probability $p_* = p({\bm r}^N_*)$.
 	\item Accept ${\bm r}^N_*$ as ${\bm r}^N(s+1)$ with a probability of $ \min\left( 1, p_*/p_{pre} \right) $.
 	\item If accepted $s\leftarrow s+1$, $p_{pre} \leftarrow p_*$.
\end{enumerate}
Through this process ${\bm r}^N(s)$ follows $p({\bm r}^N)$.

\subsection{Molecular Flow Metropolis-Hastings (MFMH) }
The main idea behind the MFMH method is to avoid high energy states in the proposal distribution. We transform the proposal distribution $q({\bm r}_{i*}|{\bm r}_i)$(Eq.\ref{eq41}) into $q'({\bm r}_{i*}|{\bm r}_i)$ using the inverse square flow. Such proposal distribution $q'$ generates no states which have extremely high energy, therefore, the acceptance ratio may be improved with smaller correlation. Furthermore, if the distribution $q$ is symmetric, the distribution of $q'$ is also symmetric. This is because the instantaneous change is always space-constant through transformation and therefore the ratio of $q$ to $q'$ is also constant. Thus, the calculation of the probability of $q'$ is not required.

The procedure of MFMH is as follows:
\begin{enumerate}
	\item Set sample steps $s = 0$.
	\item Set initial molecule positions before transform ${\bm h}^N(s)$.
	\item Perform transformation into ${\bm r}^N(s=0)$ from ${\bm h}^N(s=0)$ using inverse square flow.
	\item Calculate probability $p_{pre} = p({\bm r}^N(s=0))$.
	\item Repeat the process process below until a sufficient number of states have been accepted.
 	\item Generate state ${\bm h}^N_*$ from the proposal distribution $q$ and compare to the previously accepted state ${\bm h}^N(s)$.
 	\item Perform transformation into ${\bm r}^N_*$ from ${\bm h}^N_*$ using inverse square flow.
 	\item Calculate the probability $p_* = p({\bm r}^N_*)$.
 	\item Accept ${\bm r}^N_*$ as ${\bm r}^N(s+1)$ and ${\bm h}^N_*$ as ${\bm h}^N(s+1)$ with a probability of $ \min\left( 1, p_*/p_{pre} \right) $.
 	\item If accepted $s\leftarrow s+1$, $p_{pre} \leftarrow p_*$.
\end{enumerate}
Note that, for MFMH, the integration time, $T$, for performing transformations must be same among all states. By following this process we can ensure that ${\bm r}^N(s)$ follows $p({\bm r}^N)$.

\begin{figure}[tbp]
\begin{center}
\includegraphics[width=80 mm ]{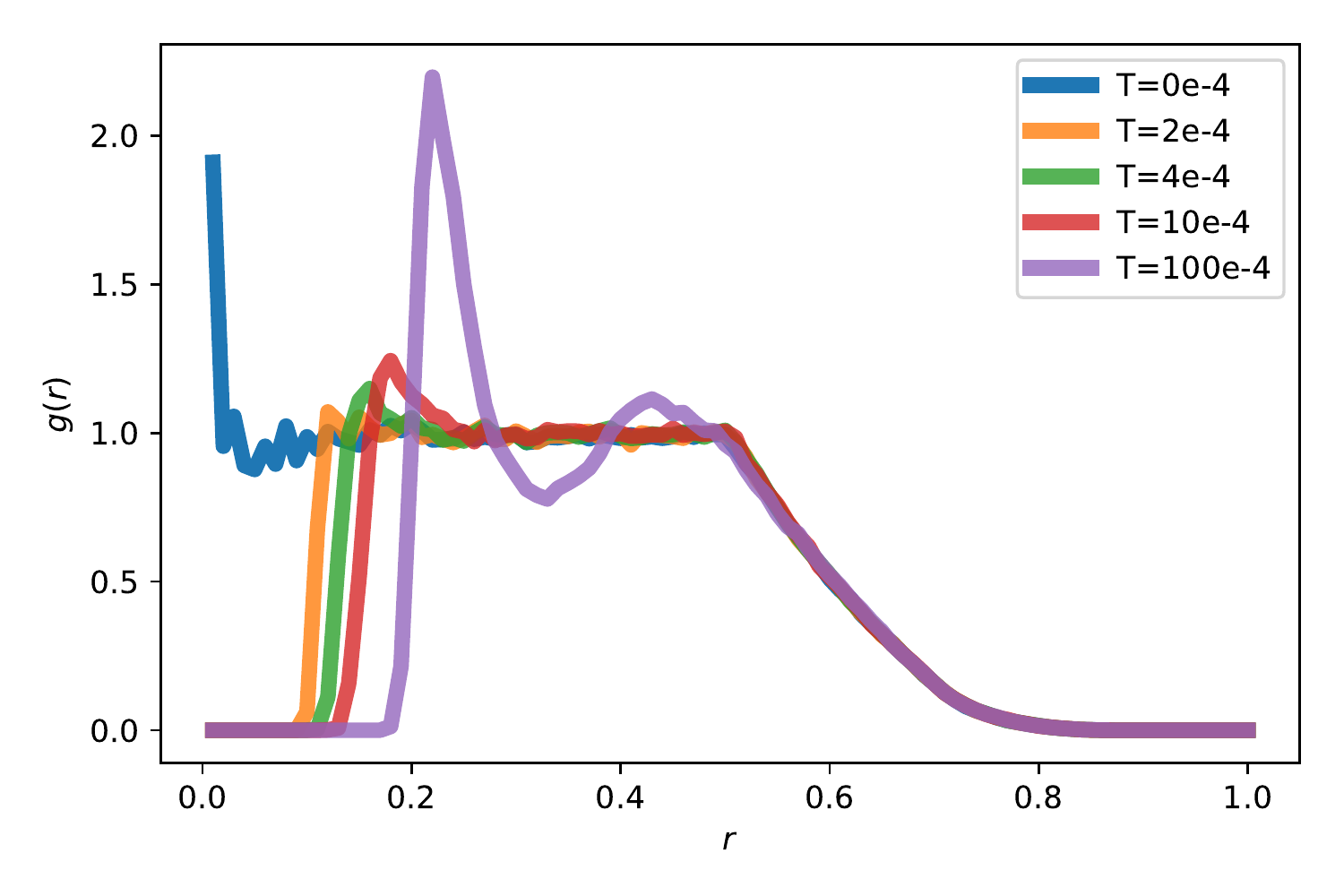}
\caption{
{\bf Radial distribution function (RDF) of a system state generated using 3-dimensional inverse square flow from a uniform distribution with periodic boundary conditions.} 
Molecules in the system with periodic boundaries were remapped using inverse square flow. 
Each line represents the RDF of a state in which molecules have been remapped using inverse square flow from a state where molecules were positioned with uniform-randomness. 
The integration time, $T$, is denoted by line colour, with greater times resulting in larger molecule-pair separation distances.}
\label{fig51}
\end{center}
\end{figure}

\section{5. Simulation results}

\subsection{Transformation of uniformly distributed molecules }
In the first test of our method, we begin by verifying the results for transformations by the inverse square flow in the presence of periodic boundary conditions. As seen in Fig.\ref{fig31}, systems that have several molecules pairs in close proximity are transformed into systems that exclude these molecule pairs. To demonstrate this clearly, we performed inverse square flow transformations on states of uniformly distributed molecules, and calculated the radial distribution functions (RDF), comparing the initial and transformed configurations. The RDF $g(r)$ describes the probability of finding a particle pair at a separation distance of $r$, relative to that of an ideal gas. The RDF $g(r)$ is defined mathematically as
\begin{eqnarray}
g(r) = \frac {\left< n(r) \right>}{ 4\pi r^2 dr \rho },
\end{eqnarray}
where $\left< n(r) \right>$ denotes the average number of particle-pairs with separation distances between $r$ and $r+dr$, $\rho$ is average density of the system.
The configuration of molecules ${\bm r}^N$ in the system before the transformation was sampled from a uniform distribution. The results of these tests are shown in Fig.\ref{fig51}.
The results in Fig.\ref{fig51} indicates that the RDF profiles for the transformed states vary with integration time, $T$. 
The blue line ($T=0$) represents the RDF of initial configuration (uniformly distributed molecules).
As time $T$ becomes longer, the separation distance between molecules becomes larger and molecule pairs in close proximity, which are the cause of extremely high energy states, are completely excluded. Therefore, these transformed states are suitable for efficient sampling.

\begin{figure}[tbp]
\begin{center}
\includegraphics[width=80 mm ]{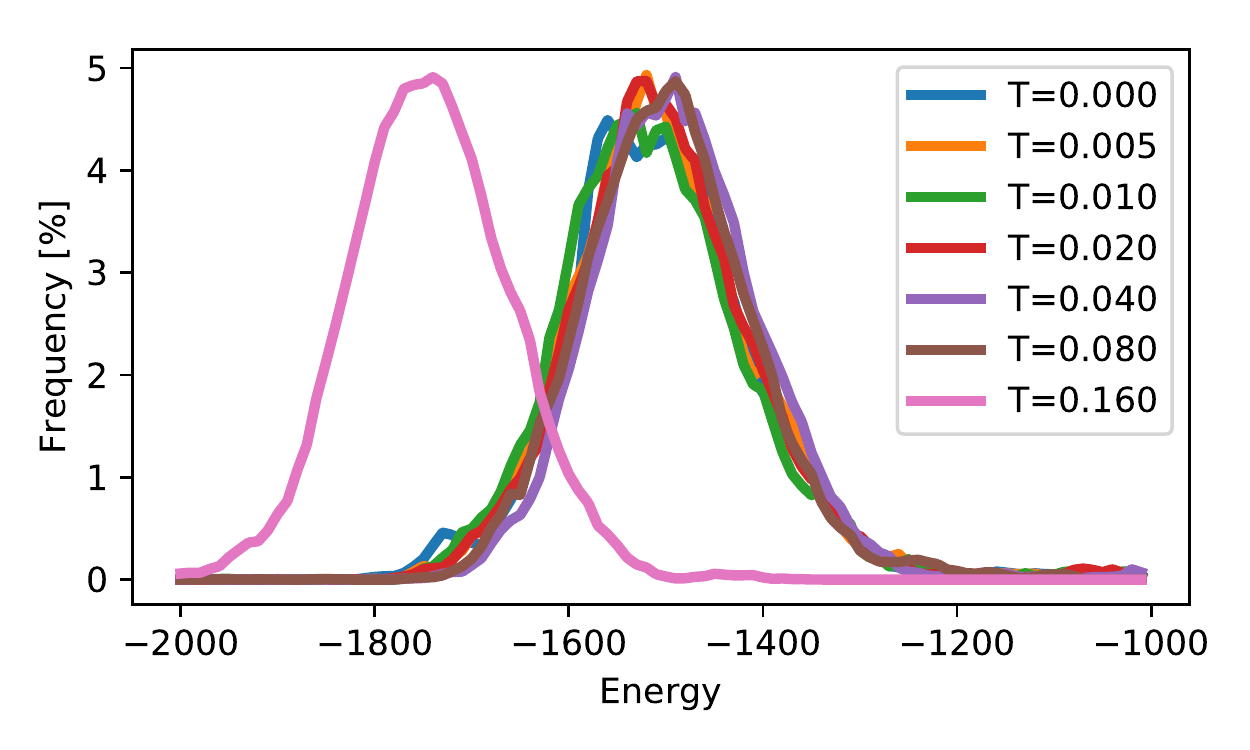}
\caption{
{\bf Energy distribution of LJ ensembles using standard MH and MFMH.} 
LJ systems, simulated in the NVT ensemble ($Temperature$ = 4 and $\rho$ = 1), are sampled with standard MH and MFMH methods. 
Each line represents energy distributions for states sampled using the standard MH method ( $T=0.000$ ) and the MFMH method ( $T=0.005, 0.010, 0.020, 0.040, 0.080, 0.160$ ). $T$ is integration time of MFMH.}
%All lines, apart from $T=0.160$, have the same energy distribution. These matching distributions indicate that MFMH methods correctly sample the ensembles. The reason that the wrong distribution was obtained for the case of $T=0.160$ is that MF avoids extremely low energy states which are sometimes necessary for correct MH sampling.}
\label{fig52}
\end{center}
\end{figure}

\begin{figure}[tbp]
\begin{center}
\includegraphics[width=80 mm ]{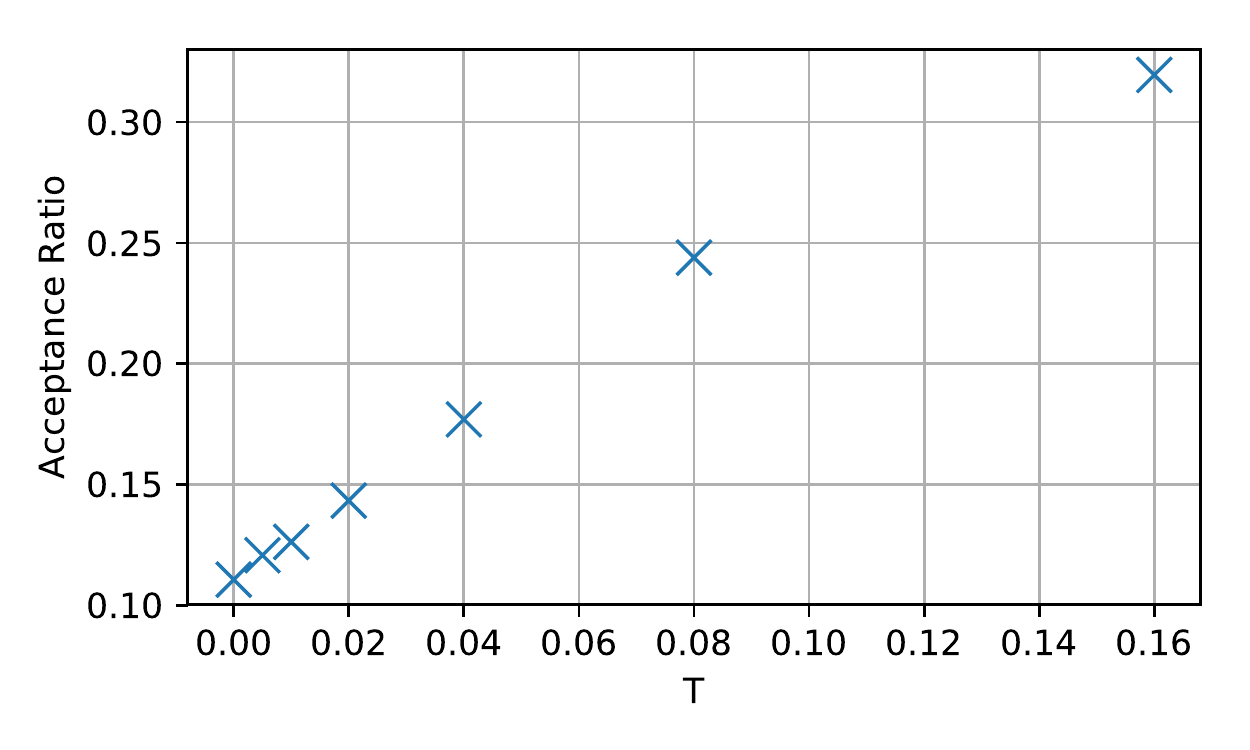}
\caption{
{\bf Acceptance ratio for LJ systems simulated in the $NVT$ ensemble with standard MH and MFMH methods. } 
LJ systems, simulated in the NVT ensemble ($Temperature$ = 4 and $\rho$ = 1), are sampled with standard MH and MFMH methods. The line at $T = 0.00$ is the acceptance ratio of states using standard MH method, while the other lines were simulated using MFMH. $T$ denotes integration time of MFMH.}
%Aside from the case of $T=0.160$, which does not give the correct distribution ( Fig.\ref{fig52} ), it was found that, in the best case, the acceptance ratio when using the MFMH method is more than twice that of the standard MH method.}
\label{fig53}
\end{center}
\end{figure}

\subsection{Sampling Lennard-Jones particles system by MFMH method}
Next, we confirmed the efficiency of the MFMH method using by simulation of a Lennard-Jones(LJ) particle system. This systems contains only monoatomic molecules whose interactions depend only on the LJ potential \cite{Jones1924}. For LJ particles, the two-body intermolecular potential energy $U_{ij}(r)$ is represented in dimensionless units as:
\begin{eqnarray}
U_{ij}(r) = 4\left( r^{-12} - r^{-6} \right),
\end{eqnarray}
where $r = \| {\bm r}_i - {\bm r}_j \|$.
To test the efficiency of the MFMH method, we compared the acceptance ratios between the standard MH and the MFMH methods for the same proposal distribution $q$ (Eq.\ref{eq41}) and molecule movement range ${r_{move}}$. The target system contains $N = 64$ LJ particles, and the system $Temperature$ and particle density $\rho$ are 4 and 1, respectively.
%time scaling is not mentioned.
The results of these simulations are shown in Fig.\ref{fig52} and Fig.\ref{fig53}. We generated $10^6$ proposal states using the standard MH method, and $10^6$ proposal states using the MFMH method. The molecule movement range was set to ${r_{move}}=0.05$. As a result, the energy distribution of states for both methods showed good agreement, except for the case where $T=0.160$. Further to this, the acceptance ratios were found to be 0.112 for the standard MH method and 0.244 for the MFMH method at $T=0.080$. Consistency of the energy distributions indicates that the MFMH method accurately sampled the system ensemble. These energy distributions were also consistent with previous research \cite{Karl1993}.
The reason that the case for $T=0.160$ produces in an incorrect distribution can be considered a result of the fact that MF avoids excessively low energy states, which are necessary for correct MH sampling.
Furthermore, even though the same value of ${r_{move}}$ was used, the acceptance ratio for the MFMH method is, in the best case, more than twice that of the standard MH. It may be claimed with some justification, therefore, that the MFMH method samples states more efficiently, and with the same level of accuracy, as the standard MH method.

\section{6. Conclusions}
In this work, we have proposed a novel and efficient sampling method ``Molecular Flow Monte Carlo'' using inverse square flow. Inverse square flow transforms any distribution into a new distribution which precludes the possibility of molecule pairs occurring in extremely close proximity, with a constant change in probability density. Therefore, we can avoid high energy states from the proposal distribution of the Metropolis-Hasting method, which improves efficiency. This improvement of efficiency has been confirmed through case study simulations of molecular systems.

\bibliographystyle{aaai.bst}
\bibliography{citation}

\begin{thebibliography}{}

\bibitem[\protect\citeauthoryear{Andrzej and Skolnick}{1994}]{Andrzej1994}
Andrzej, K., and Skolnick, J.
\newblock 1994.
\newblock Monte carlo simulations of protein folding. i. lattice model and
  interaction scheme.
\newblock {\em Protein. Struct. Funct. Bioinformat.} 18(4):338--352.

\bibitem[\protect\citeauthoryear{Chen \bgroup et al\mbox.\egroup
  }{2018}]{Ricky2016}
Chen, R. T.~Q.; Rubanova, Y.; Bettencourt, J.; and Duvenaud, D.
\newblock 2018.
\newblock Neural ordinary differential equations.
\newblock {\em arXiv preprint arXiv:1806.07366}.

\bibitem[\protect\citeauthoryear{Escobedo and de Pablo}{1996}]{Escobedo1996}
Escobedo, F.~A., and de~Pablo, J.~J.
\newblock 1996.
\newblock Expanded grand canonical and gibbs ensemble monte carlo simulation of
  polymers.
\newblock {\em J. Chem. Phys.} 105(10):4391--4394.

\bibitem[\protect\citeauthoryear{Ewald}{1921}]{Ewald1921}
Ewald, P.~P.
\newblock 1921.
\newblock Die berechnung optischer und elektrostatischer gitterpotentiale.
\newblock {\em Ann. Phys.} 369(3):253--287.

\bibitem[\protect\citeauthoryear{Hastings}{1970}]{Hastings1970}
Hastings, W.~K.
\newblock 1970.
\newblock Monte carlo sampling methods using markov chains and their
  applications.
\newblock {\em Biometrika} 57(1):97--109.

\bibitem[\protect\citeauthoryear{Jones and Chapman}{1924}]{Jones1924}
Jones, J.~E., and Chapman, S.
\newblock 1924.
\newblock On the determination of molecular fields. -ii. from the equation of
  state of a gas.
\newblock {\em Proc. R. Soc. Lond. A} 106(738):463--477.

\bibitem[\protect\citeauthoryear{Karl, Zollweg, and Gubbins}{1993}]{Karl1993}
Karl, J.~J.; Zollweg, J.~A.; and Gubbins, K.~E.
\newblock 1993.
\newblock The lennard-jones equation of state revisited.
\newblock {\em Mol. Phys.} 78(3):591--618.

\bibitem[\protect\citeauthoryear{Mandell and McTague}{1976}]{Mandell1976}
Mandell, M.~J., and McTague, J.~P.
\newblock 1976.
\newblock Crystal nucleation in a three-dimensional lennard-jones system: A
  molecular dynamics study.
\newblock {\em J. Chem. Phys.} 64(9):3699--3702.

\bibitem[\protect\citeauthoryear{Mezei \bgroup et al\mbox.\egroup
  }{1983}]{Mezei1983}
Mezei, M.; Beveridge, D.~L.; Berman, H.~M.; Goodfellow, J.~M.; Finney, J.~L.;
  and Neidle, S.
\newblock 1983.
\newblock Monte carlo studies on water in the dcpg/proflavin crystal hydrate.
\newblock {\em J. Biomol. Struct. Dyn.} 1(1):287--297.

\bibitem[\protect\citeauthoryear{Sugita and Okamoto}{1999}]{Sugita1999}
Sugita, Y., and Okamoto, Y.
\newblock 1999.
\newblock Replica-exchange molecular dynamics method for protein folding.
\newblock {\em Chem. Phys. Lett.} 314(1):141--151.

\bibitem[\protect\citeauthoryear{Torrie and Valleau}{1977}]{Torrie1977}
Torrie, G.~M., and Valleau, J.~P.
\newblock 1977.
\newblock Nonphysical sampling distributions in monte carlo free-energy
  estimation: Umbrella sampling.
\newblock {\em J. Comp. Phys.} 23(2):187--199.

\bibitem[\protect\citeauthoryear{Yi, Cong, and Zhonghan}{2017}]{Yi2017}
Yi, S.; Cong, P.; and Zhonghan, H.
\newblock 2017.
\newblock Note: A pairwise form of the ewald sum for non-neutral systems.
\newblock {\em J. Chem. Phys.} 147(12):126101.

\end{thebibliography}

\section {Appendix}

\subsection{A. Non-derivative form of the pairwise force of inverse square flow}

\begin{eqnarray}
{\bm \nabla}_{\bm r} \left(
\sum_{\bm n} { \frac{{\rm erfc}(G\|{\bm r}-{\bm n}\|)}{\|{\bm r}-{\bm n}\|  }} 
+
4\pi \sum_{\bm k \neq 0}{
\frac{ e^{-\|\bm k\|^2/4G^2} e^{i{\bm k}\cdot {\bm r}}  }{\|\bm k\|^2}
}
\right)
\end{eqnarray}
may be rewritten as
\begin{eqnarray}
-\sum_{\bm n} { \frac{{\bm r}-{\bm n}}{\|{\bm r}-{\bm n}\|^3} }
	\biggl\{
	{\rm erfc}(G\|{\bm r}-{\bm n}\|) \nonumber \\
	\left. +\frac{2G}{\sqrt{\pi}} {\|{\bm r}-{\bm n}\|} e^{-G^2 {\|{\bm r}-{\bm n}\|^2} }
	\right\}
 \\
-4\pi \sum_{\bm k \neq 0}{
	\frac{ e^{-\|\bm k\|^2/4G^2} }{\|\bm k\|^2} {\bm k} \sin{({\bm k}\cdot {\bm r})},
}
\end{eqnarray}
by performing differentiation.

\subsection{B. Instantaneous change of the periodic sum of inverse square flow}
We now investigate the instantaneous change of the following flow,
\begin{eqnarray}
  		\frac{d{\bm r}_i}{dt}= \sum_{i \neq j}{ {\bm f}^p({\bm r}_i-{\bm r}_j) }.
\end{eqnarray}
${\bm f}^p({\bm r})$ is constructed from a real-space term and reciprocal-space term. The real-space term is
\begin{eqnarray}
{\bm \nabla}_{\bm r}
\left(
	\sum_{\bm n} {   \frac{{\rm erfc}(G\|{\bm r}-{\bm n}\|)}{\|{\bm r}-{\bm n}\|  }}
\right).
\label{eqB1}
\end{eqnarray}
Incidentally, if $f_{ij}(r){\bm r}$ is represented as ${\bm \nabla}_{\bm r}h_{ij}(r)$ by using some function $h_{ij}$, Eq.\ref{eq11} can be expressed as
\begin{eqnarray}
		{\rm tr}\left( \frac{df}{d{\bm r}^N(t)} \right) = 
		\sum_{i\neq j}  \frac{2}{r_{ij}} \frac{dh_{ij}}{dr_{ij}} + \frac{d^2h_{ij}}{{dr_{ij}}^2} 
\label{eqB2},
\end{eqnarray}
for $D=3$. Using Eq.\ref{eqB2}, and assuming ${\bm f}^p({\bm r})$ has only real-space terms, the instantaneous change may be calculated as
\begin{eqnarray}
		{\rm tr}\left( \frac{df}{d{\bm r}^N(t)} \right) = 
		\sum_{i\neq j}  \frac{4G^2}{\sqrt{\pi}}
		\sum_{\bm n}{
		e^{-G^2 \| {\bm r}_i-{\bm r}_j-{\bm n} \|^2   } 
		}
\end{eqnarray}
by setting 
\begin{eqnarray}
h_{ij}(r)=\sum_{\bm n} \frac{{\rm erfc}(G\|{\bm r}-{\bm n}\|)}{\|{\bm r}-{\bm n}\|  }.
\end{eqnarray}
If we now think about the screening factor $G$ in the limit of $G \rightarrow 0$, the reciplocal-space term goes to $0$, therefore
\begin{eqnarray}
		{\rm tr}\left( \frac{df}{d{\bm r}^N(t)} \right) = 
		\sum_{i\neq j} \lim_{G \rightarrow 0} \frac{4G^2}{\sqrt{\pi}}
		\sum_{\bm n}{
		e^{-G^2 \| {\bm r}_i-{\bm r}_j-{\bm n} \|^2  } 
		} \nonumber \\
		=
		\sum_{i\neq j} \frac{4}{\sqrt{\pi}}
		\int_{ {\bm x} \in \mathbb{R}^3 }
		e^{- \|{\bm x}\|^2   } \nonumber 
		\\
		= \sum_{i\neq j} \frac{4}{\sqrt{\pi}} {\pi^{3/2}} =  \sum_{i\neq j} 4\pi
\end{eqnarray}
holds. This equation holds any value of $G \in (0,\infty)$ because the screening factor $G$ only divides the inverse square flow into real and reciprocal terms and their sum is constant.

\end{document}